\RequirePackage{fix-cm}
\documentclass{svjour3}                     % onecolumn (standard format)
\smartqed  % flush right qed marks, e.g. at end of proof
\usepackage{graphicx}
\usepackage{amsmath}
\usepackage{algorithm}
\usepackage{algpseudocode}
\usepackage{algorithmicx}
\usepackage{hyperref}
\usepackage{tabularx}
\usepackage{color}
\begin{document}

%JOCHEN: First, I asked myself whether the current title describes what you are doing in the paper. Would it be possible to have a reference to clustering solutions in there? Could the subtitle (Browsing through the universe of bibliographic information) be a bit more specific?

\title{Mutual Information based labelling and comparing clusters}

%\subtitle{Browsing through the universe of bibliographic information}
%ASOct14 title and subtitle could also be inverted

\author{Rob Koopman         \and
        Shenghui Wang 
}

\institute{R. Koopman \at
              OCLC Research, Schipholweg 99, Leiden, The Netherlands \\
              Tel.: +31 71 524 6500\\
              \email{rob.koopman@oclc.org}           %  \\
           \and
           S. Wang \at
              OCLC Research, Schipholweg 99, Leiden, The Netherlands \\
              Tel.: +31 71 524 6500\\
              \email{shenghui.wang@oclc.org}           %  \\
% 			\and 
%             A. Scharnhorst \at
%               DANS-KNAW, Anna van Saksenlaan 51, The Hague, The Netherlands \\
%               Tel.: +31 70 349 4450\\
%               \email{andrea.scharnhorst@dans.knaw.nl}           %  \\
% %             \emph{Present address:} of F. Author  %  if needed
}

\date{Received: date / Accepted: date}

\maketitle

\begin{abstract}
After a clustering solution is generated automatically, labelling these clusters becomes important to help understanding the results. In this paper, we propose to use a Mutual Information based method to label clusters of journal articles. Topical terms which have the highest Normalised Mutual Information (NMI) with a certain cluster are selected to be the labels of the cluster. Discussion of the labelling technique with a domain expert was used as a check that the labels are discriminating not only lexical-wise but also semantically. Based on a common set of topical terms, we also propose to generate \textit{lexical fingerprints} as a representation of individual clusters. Eventually, we visualise and compare these fingerprints of different clusters from either one clustering solution or different ones. 

\end{abstract}

\section{Introduction}
\label{sec.intro}
%It is not straightforward to pick descriptive, human-understandable labels to summarize the content of the clusters produced by an automated clustering algorithm. Labelling clusters are often done by those producing them on the basic of common or specific tacit knowledge. Another approach it to apply techniques which compare term distributions across clusters, such as mutual information, and use them for labelling clusters~\cite{IR2008}. In the following we apply mutual information to identify topical terms with a potential of being a label. We also used mutual information to generate a list of discriminatory topical terms for all clusters and used those to visually compare clusters.

Identifying thematic structures in science (so-called topics) is the shared goal for all the different methods described in this special issue ``Same data, different results?'' Every method produced a set of clusters, with each cluster grouping similar or relevant articles together, to reflect certain thematic structures in the same dataset. Comparing different clustering solutions is however a challenge in itself, as reported in~\cite{velden2015comparison}. What ever numeric measures we apply, such as shared documents across different solutions, or size distributions, those measures give little insight into the meaning of the differences between clustering solutions. For the interpretation of clustering solutions and to relate them to the research fields that people know about, it seems more natural and intuitive if we could assign human-understandable labels to describe the content of those clusters or the \textit{topics} they represent. This paper presents a method to first assign labels and second to compare them at a more abstract yet still meaningful level. Although, the approach has been developed as a part of the ``Same data, different results?'' collaboration, we believe that it could also be applied in clustering of other objects. 

It is not straightforward to pick descriptive, human-understandable labels to summarize the content of clusters produced by an automated clustering algorithm~\cite{IR2008}. Labelling clusters is often done by those producing them on the basic of common or specific tacit knowledge. When doing automatic clustering of documents, which also contain lexical information in titles, keywords, abstracts, publication venues and alike, using measures based on frequency and co-occurrence of terms come to mind. For example, one could look at the most frequent terms in the bibliographic metadata of the documents belonging to a cluster. Or, one could consider to use terms that occur frequently in the centroid (the middle of a cluster)
%ASFeb24 here it is not clear what you mean by centroid, so another sentence is needed to introduce the centroid
or the documents that lies closest to the centroid. 

For this paper we follow another labelling approach, namely to use measures such as mutual information to compare distributions of terms in one cluster with that of other clusters. Those terms are extracted from the lexical information of the documents, in our case, the titles and abstracts of the articles. We call this a \textit{differential cluster labelling} approach, because it selects those terms which are frequent in one cluster but are not frequent in others as potential labels for this cluster. The advantage of this method is that it is independent of the clustering method, because it only uses the terms extracted from the articles' title and abstract against the final cluster assignments. Furthermore, it can be applied independently of the availability of assigned keywords or subject headings which are commonly used. 

%ASFeb24: I think somewhere here you need to say which sources you use for the terms, where do they come from. In the next section you talk about comparison of terms and articles. A reader not familiar with the whole work could think we talk about terms from the whole text of the article but this is not true! 
%Also here, and in the abstract it is not clear what the stepwise process is. Creating 'significant terms' from the whole ensemble of articles, for one clustering solution - which one? From those significant terms youd evelop labels, right? and how you go than from one set of labels to compare them across solutions. You explained this very clearly in the presentation in the eHg; but I have to admit that I forgot this.

Having labels based on significant terms which are human-understandable should contribute to the comparison of clustering solutions. %, if a common label-based coordinators could be determined. 
%We first apply the technique to label individual clusters.  
%Feb24 see commment above, all clusters individually , or what?
We further extend the labelling approach and identify a set of terms which are most informative for all clustering solutions that we need to compare. We then generate a \textit{fingerprint} for each cluster by measuring their Normalised Mutual Information against these labels. %We have however at least eight different clusterings to compare. Therefore we felt the need to have a visual for comparing clusters. To achieve this we used mutual information to generate a list of fifty discriminatory topical terms for each clustering method. We combined the lists and removed semantically and textual near duplicates resulting in a final list of 61 terms. After calculating the NMI between each cluster and each term of the list the result is a vector of 61 values for each cluster. 
We order these selected terms as solving the Travelling Salesman Problem~\cite{tsp}, resulting a one-dimensional word-space, where each term has a specific coordinate.  
%Feb24AS: maybe I create confusion by differentiating between terms and labels, get rid of this if this is the case, you order the labeling terms according to xxxx, it was not alphabetic, but what?
Now we can visualize the fingerprints of all clusters in terms of the Normalised Mutual Information and compare them using those common \textit{label-based} coordinates. This gives us direct insight in the qualitative differences between clusters across different clustering solutions. 

The main goal of this paper is to apply methods based on Normalised Mutual Information to label and compare clusters. In the first part  we describe our experiment of using the Normalised Mutual Information to identify labels for individual clusters from different clustering solutions.
%ASFeb24: again for which set of clusters?
To test the meaningfulness of the labels we discussed them with a domain expert. 
%ASFeb24: Make sure that Marcus agrees to the use of the figure, I saw that you thanks him in the acknowledgement. 
The second part of the paper is about comparing clusters based on their label-based fingerprints. 

\section{Mutual information based labelling}

%AS @Andrea from the old text, add that labeling is always done with a lot of tacit knowledge, also maybe some of the ISSI insights can be reproduced; we can get some insights from the terms which pop up most; but this method has been seen as not metric enough; and also the iterative process - to go to Wikipedia, Scholar, and a like might help us to gain insights but does not really delivers words we feel comfortable to use as labels
%by the way: the comparison article uses a lot of labels - without - at the moment syaing how those come about
%ASFeb24: Tacit knwoledge ;-)

Mutual Information is a common technique for labelling  clusters~\cite{IR2008}. 
%Once articles are clustered into different clusters, we measure the Normalized Mutual Information (NMI) between topical terms and clusters. 
The Mutual Information measures how much information the presence/absence of a term $t$ contributes to making the correct clustering decision on a cluster $c$. In other words, the Mutual Information represents the reduction in uncertainty about the cluster $c$ given the knowledge of the term $t$. A high Mutual Information
%@Shenghui: make sure you always use capital initials for Mutual Information through the text or even not - the text above is very clear!
between a term $t$ and a cluster $c$ suggests this term describes a large part of the content of this cluster therefore this term could be a candidate for labelling this cluster. 

Formally, the mutual information between a term $t$ and a cluster $c$ is calculated as follows:
\begin{equation}
I(t,c) = \sum_{i=0}^{1} \sum_{j=0}^{1} P(T_i, U_j) \log_2 \frac{P(T_i, U_j)}{P(T_i) P(U_j)}
\label{eq.mi}
\end{equation} 
where $T_0$ indicates an article does not contain the term $t$ and otherwise $T_1$, $U_0$ indicates an article does not belong to the cluster $c$ and otherwise $U_1$, $P(T_i, U_j)$ is the probability that $T_i$ and $U_j$ happen together within one article, $P(T_i)$ and $P(U_j)$ are the probabilities that these events happen independently. The probabilities are estimated by dividing the frequency of the observed event (the article contains or does not contain the term $t$ or it is in or not in the cluster $c$) by the total number of articles. 

We then normalize this mutual information $I(t,c)$ by dividing over the entropy of cluster $c$, i.e.,
\begin{equation}
\label{eq.nmi}
NMI(t,c)=2 \times \frac{I(t,c)}{H(t)+H(c)}
\end{equation}
where 
\begin{equation}
H(t)= -\sum_{i=0}^1 P(T_i) \log_2 P(T_i) \textrm{ and } H(c)=-\sum_{j=0}^1 P(U_j) \log_2 P(U_j).
\end{equation}

This Normalised Mutual Information (NMI) score is non-negative. When a term and a cluster are completely independent from each other, the NMI score is 0. When a term only occurs frequently in one cluster but rarely in others, then the NMI score between this term and this cluster is high. When a relatively frequent term has an extraordinarily low occurrence in one cluster, the NMI score between this term and the cluster is also pretty high, indicating that the cluster is not about this term. In order to distinguish the positive and negative associations between terms and cluster, we assign a negative sign to those NMI scores when the occurrence of a term is less than expected.\footnote{If a cluster covers 10\% of the total dataset, then the term is expected to occur 10\% of its occurrences over the total dataset.}    

For each cluster in a specific clustering solution, we calculate the NMI scores between this cluster and all the 60 thousand topical terms,\footnote{The topical terms were extracted from the titles and abstracts of the articles in the \textit{Astro} dataset. Please refer to \cite{koopman2015_ariadne} for more details.}
%ASFeb24 here you need to explain what you mean by "all the topical terms" , you mean the 600 right? cite the main article
and rank these terms based on their NMI scores. The terms which have the highest NMI scores are good candidates for labelling this particular cluster. 

Table~\ref{tab.labels} shows the  topical terms with top 10 highest NMI scores for our K-Means clusters~\cite{koopman2015_clustering}. The same method was used to label all the clusters from different solutions described in this special issue. 
As shown here, this method is independent of how these clusters are generated, as it only uses the information from the articles which are in the clusters. Also these topical terms are extracted automatically from the titles and abstracts of all articles~\cite{koopman2015_ariadne}. As said before, this labelling method does not depend on the availability of the pre-assigned keywords or subject headings. It is actually generalisable to label any collections of articles. Because it is a data-driven approach based on lexical information, these topical terms are sometimes only understandable to domain experts, and most likely part of the very specific vocabulary in this domain. Our impression, alternating between articles and the identified labels, was that the labels do reflect potential topics, and do so on a lower level of abstraction or more specific than  keywords chosen by authors or librarians.

%ASFeb24: I feel like you need another summarizing sentence here. Will you have an appendix with all suggested labels for all clusters, or would this take too much spacea and/or be too redundant?
%ASFeb24: Also in the tables you show the 10 most significant labels, but for the visualisations you use more, say something about this.

\begin{table}
\centering
\caption{K-Means cluster labels
\label{tab.labels}}
\small{
\begin{tabularx}{\textwidth}{ccX}\hline
Cluster ID & Size& Cluster labels\\\hline
ok 0 & 2866&seyfert 1, active galactic, narrow line, agn, broad line, galactic nuclei, quasars, line seyfert, nuclei agns, emission line\\
ok 1&2934&lens, microlensing, gravitational lens, rotation curve, spiral galaxies, bars, dark matter, barred galaxies, galaxy, pattern speed\\
ok 2&5449&transit, star, eclipsing binary, radial velocity, hd, planet, corot, photometric, main sequence, type stars\\
ok 3&3389&spacetimes, black hole, horizon, asymptotically flat, reissner nordstrom, metric, einstein maxwell, spherically symmetric, hole solutions, schwarzschild\\
ok 4&5420&solar wind, magnetosphere, interplanetary magnetic, magnetic field, auroral, plasma, magnetopause, ion, substorm, spacecraft\\
ok 5&3874&standard model, higgs, lhc, minimal supersymmetric, supersymmetric standard, neutrino mass, lepton, right handed, hadron collider, electroweak\\
ok 6&4721&quark, qcd, meson, decays, lattice qcd, pi pi, pion, j psi, form factors, chiral\\
ok 7&4206&galaxy clusters, dark matter, haloes, cluster, n body, weak lensing, intracluster medium, halo mass, 1 mpc, galaxies\\
ok 8&2195&blazar, bl lac, jet, radio sources, lac objects, radio galaxies, synchrotron, radio, flat spectrum, 3c\\
ok 9&3225&yang mills, gauge theory, mills theory, string, supergravity, noncommutative, field theory, supersymmetric, dual, branes\\
ok 10&4070&globular clusters, fe h, metal poor, red giant, metallicity, giant branch, horizontal branch, galactic globular, color magnitude, stars\\
ok 11&3409&ray binary, x ray, hard state, ray timing, rossi x, timing explorer, black hole, accretion disk, neutron star, rxte\\
ok 12&6022&galaxies, star formation, formation rate, deep field, redshift, early type, sample, rest frame, starburst, lyman break\\
ok 13&5556&quantum gravity, quantum, loop quantum, spacetime, general relativity, scalar field, gravity, quantum cosmology, metric, quantization\\
ok 14&5583&coronal, active region, solar, flare, magnetic flux, cme, quiet sun, chromosphere, mass ejections, hinode\\
ok 15&2569&cosmic ray, high energy, gamma rays, tev, hess, ultra high, air showers, tev gamma, extensive air, shower\\
ok 16&1985&microwave background, cosmic microwave, background cmb, cmb, microwave anisotropy, anisotropy probe, wilkinson microwave, wmap, power spectrum, probe wmap\\
ok 17&2465&dark energy, quintessence, universe, phantom, f r, cosmological constant, cosmic acceleration, chaplygin gas, modified gravity, accelerated expansion\\
ok 18&2206&white dwarf, cataclysmic variables, dwarf nova, nova, wd, mass transfer, orbital period, secondary star, cvs, superhumps\\
ok 19&1627&inflation, slow roll, curvature perturbation, non gaussianity, inflationary models, curvaton, reheating, cosmological perturbations, f nl, primordial\\
ok 20&2020&gravitational wave, inspiral, ligo, lisa, wave detectors, laser interferometer, waveforms, binary black, space antenna, post newtonian\\
ok 21&1849&grb, ray burst, gamma ray, afterglow, bursts grbs, swift, prompt emission, prompt, fireball, batse\\
ok 22&4103&asteroid, comet, body problem, orbits, kuiper belt, main belt, bodies, mean motion, planets, solar system\\
ok 23&2071&planetary nebulae, asymptotic giant, agb stars, post agb, giant branch, pne, branch agb, agb, central star, mira\\
ok 24&4592&molecular cloud, protostellar, cloud, c 13, star forming, h 2, molecules, hco, forming regions, massive star\\
ok 25&2622&ionospheric, winter, summer, degrees n, mesosphere, tec, electron content, iri, ozone, seasonal\\
ok 26&3593&mars, titan, ice, water, deposits, cassini, co2, methane, atmosphere, surface\\
ok 27&6292&performance, scientific, technology, mission, astronomical, development, research, flight, cost, software\\
ok 28&4903&sn, explosion, wolf rayet, type ia, supernova, ejecta, wr, progenitor, eta carinae, lines\\
ok 29&3176&brown dwarfs, tauri stars, pre main, herbig ae, substellar, circumstellar disks, young, main sequence, disks, low mass\\
ok 30&2624&pulsar, neutron stars, psr, radio pulsars, anomalous x, magnetar, isolated neutron, soft gamma, millisecond pulsars, axp\\
\hline
\end{tabularx}}
\end{table}

\begin{figure}
\centering
\includegraphics[width=\linewidth]{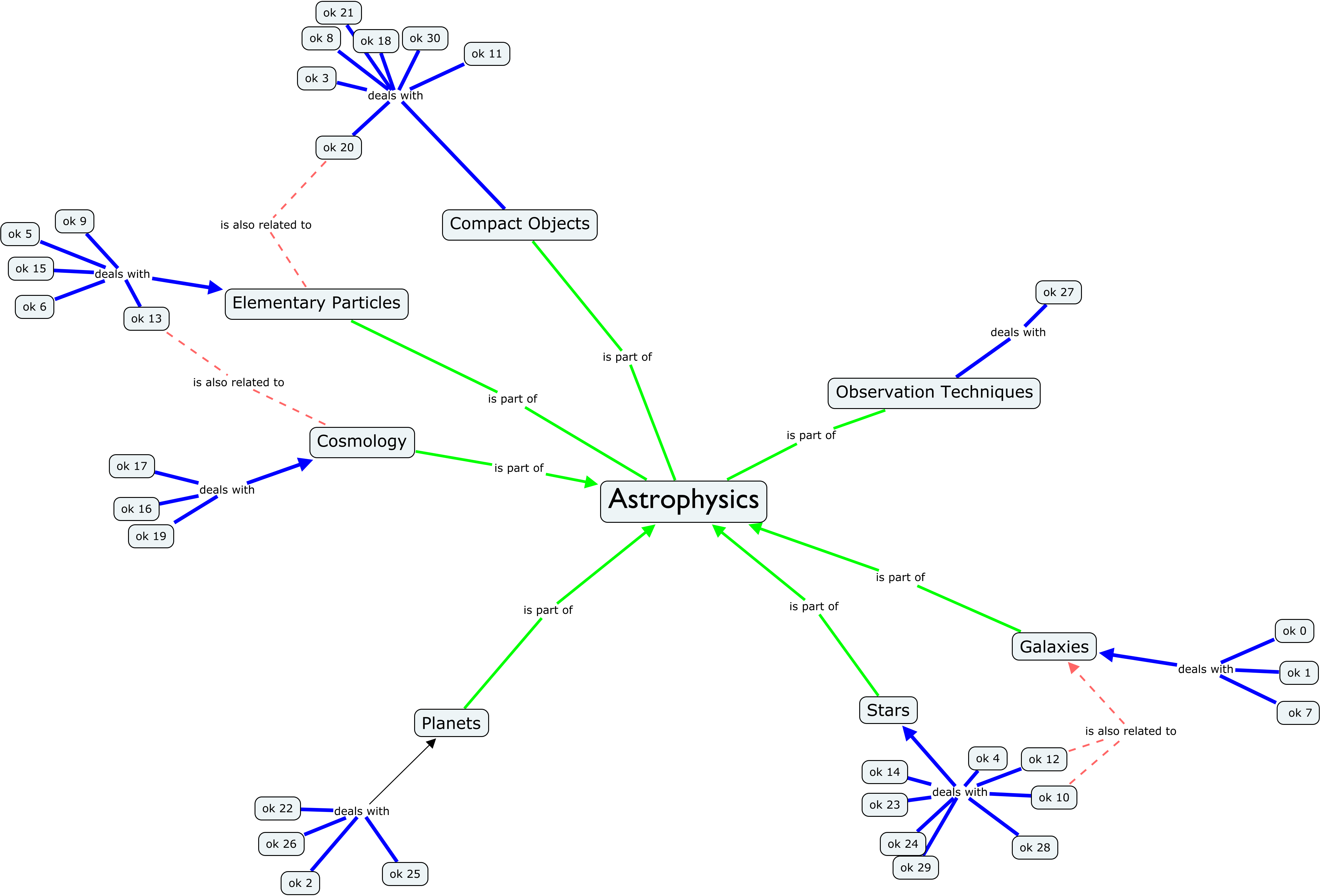}
\caption{Expert's concept map again the K-Means clusters (courtesy of Marcus John) \label{fig.conceptmap}}
\end{figure}

\paragraph{Analysis with an domain expert} 
With the produced labels for the clusters we can now compare clusters based on those labels. Still, there is one problem, we cannot solve, there is no objective ground truth to evaluate the resulting labels. Not being experts in astrophysics, we cannot really judge how meaningful the selected labels are, compared with the content of the clusters. As cross-check, we discussed the labels for one clustering solution, shown in Table~\ref{tab.labels},  with a domain expert. The fact that this expert also has a background in bibliometrics helped in the discussion. We explained the labelling procedure and provided the domain expert with the table, without providing the articles assigned to each cluster.  In this discussion, he remarked on some inner logic between the labels and subsequently the clusters. Based on his knowledge of the Astrophysics field, he developed a concept map to order those clusters. He started with six categories of astrophysical objects at different scales: \textit{Cosmology, Galaxies, Compact Objects, Stars, Planets, Elementary Particles}, plus a general category about \textit{Observation Techniques}. He used ``is part of''  relations to draw a categorical backbone of the whole field. He then assigned each cluster to these categories and produced the concept map as shown in Figure~\ref{fig.conceptmap}. 
%Feb24AS: Did Marcus also look at the articles in the clusters or comes his concept map from the cluster labels only? You should write this.

Each cluster is connected to one of the main categories by a solid blue arrow, representing a ``deals with'' relation. This relation indicates that these clusters mostly belong to that particular category. For example, the clusters $ok 0$, $ok 1$ and $ok 7$ is all about \textit{Galaxies}. Indeed we find terms as \textit{galactic} or \textit{galaxy} among the labels for this cluster. Some clusters in the concept map also have an extra dotted red link to another categories, indicating that they are also related to the other category. For example, the cluster $ok 13$ is mostly about \textit{Elementary Particles}, but is also related to \textit{Cosmology}. This is not so much a surprise, because as we reported in~\cite{koopman2015_clustering} there are not always clear boundaries between clusters. So it is quite possible that some clusters are actually bridging two major categories. 

This exercise with one expert is not a representative evaluation. A more systematic evaluation with more experts carefully checking the articles in these clusters is of course more informative. Still this exercise gave us some confidence in the appropriateness and meaningfullness of those automatically generated labels. They were obviously specific, discriminating and at the end informative enough for a domain expert to draw a knowledge landscape about these clusters along the seven categories with good confidence. For us it was encouraging enough to trust the labels and engage in a comparison exercise between clusters from different solutions.

% %ASOct14 This could be a link to Theresa's paper and labels used there.
% %SWOCT15 Not sure about the last sentence. What are the groups of clusters? AS changed

\section{Quantitative comparing clustering solutions by labels}
Since each cluster in each solution is labelled by its most significant or informative topic terms, it is possible to find the most informative topic terms across all clusters in one solution. Those labels would than represent the whole clustering solution. Therefore, we further extend the formula (Eq.~\ref{eq.mi}) to measure the NMI scores between a whole clustering solution and all the 60K topical terms.
%ASFeb24: ok again all topical terms is which set, the set of all 600 terms? Oh, I checked my notes from the research meeting and actual there I noted 60k topical words? Needs explanation. In the equations you need to make sure about which ensemble the summation runs. It was clearer in the presentation but now reading the text, it get's blurry again - at least for me, what is different in the first round of calculation of the NMI, and in teh second ...sorry

For a clustering solution $C$ with $m$ clusters ($\{c_1, c_2, \ldots, c_m\}$), we computer  
\begin{equation}
\label{eq.mic}
I(t,C) = \sum_{k=1}^m \sum_{i=0}^{1} P(T_i, U_{k}) \log_2 \frac{P(T_i, U_{k})}{P(T_i) P(U_{k})}
\end{equation}
where $t$ is a topical term, $T_0$ indicates an article does not contain the term $t$ and otherwise $T_1$, and $U_{k}$ indicates an article belongs the cluster $c_j$. We normalise it as follows:
\begin{equation}
NMI(t,C)=2 \times \frac{I(t,C)}{ (H(t)+H(C))}
\end{equation}
where, 
\begin{equation}
H(t)=-\sum_{i=0}^{1} P(T_i) \log_2 P(T_i) \textrm{ and } H(C)=-\sum_{i=1}^m P(U_i) \log_2 P(U_i).
\end{equation}
The sign of the final NMI score is also assigned in the same way as described in the previous section. The probabilities are estimated by dividing the frequency of the observed event by the total number of articles which have a cluster assignment. This is different from labelling the individual clusters described in the previous session. The reason is that not all clustering solutions have a full coverage of the whole dataset.\footnote{Please see Table 1 in~\cite{koopman2015_ariadne}.} We only look at the part of the dataset which is covered by the clustering solution and consider that the rest of the dataset do not contribute to the information of the clusters.  
%@Rob: Does this make sense? 
%ASFeb24, In the second operation you only use the articles with have a cluster assignment, in ethe first operation you use all? Nope, don't get it....
%ASFeb24 We could ask Frank to particular check the fomulas and there explanations for consistency. 

For a clustering solution, we can now compute topic terms which have the highest  NMI scores, i.e, they contain the most information about all the clusters in this solution.  Table~\ref{tab.nmi_clustering} gives the top ten terms, in the descending order of their NMI scores, for seven clustering solutions described in this special issue. 
\begin{table}[h]
\centering
\begin{tabularx}{\textwidth}{lX}\hline
Clustering & Top ten terms \\\hline
CWTS-C5& galaxies, stars, x ray, solar, gamma ray, black hole, star formation, redshift, stellar, magnetic field, cosmological\\
UMSI0 & galaxies, stars, x ray, solar, black hole, star formation, gamma ray, stellar, redshift, observations\\
OCLC-31 & galaxies, stars, x ray, gamma ray, black hole, star formation, solar, magnetic field, dark matter, redshift\\ 
OCLC-Louvain & galaxies, x ray, stars, black hole, solar, star formation, gamma ray, redshift, cluster, similar \\
STS-RG & galaxies, stars, solar, observations, similar, x ray, magnetic field, star formation, dark matter, observed\\ 
ECOOM-BC13 & galaxies, x ray, stars, dark matter, solar, star formation, black hole, gamma ray, cosmological, magnetic field\\ 
ECOOM-NLP11 & black hole, gamma ray, galaxies, dark matter, magnetic field, stars, x ray, star formation, black holes, microwave background\\ \hline
\end{tabularx}
\caption{Top ten most informative terms for different clustering solutions \label{tab.nmi_clustering}}
\end{table}

As Table~\ref{tab.nmi_clustering} shows, the most informative terms for these clustering solutions are very similar. That means these methods actually use the similar information to make major clustering decisions, while their differences lie more in how to handle less informative terms which are not listed in this table. The order of these terms, based on their NMI score, is also similar for most of the solutions, with ``galaxies'' as the most informative term. However, ECOOM-NLP11 ranks ``black hole'' and ``gamma ray'' at the top, which is different from the others. And compared to other solutions, ``dark matter'' ranked relatively higher for the two ECOOM solutions. 

This qualitative analysis invited us to do a quantitative comparison of individual clusters based on their information content. To be able to compare all the clusters using the same coordinates, we collected the top 50 labels computed from the seven clustering solutions listed in Table~\ref{tab.nmi_clustering}. We kept all terms occurring in at least two lists of labels for clustering solutions and removed all duplicates. This leads to 61 different labels in total. In a next step we ordered these labels in a way that the sum of the distances between neighbouring labels is the minimal, i.e. similar labels are positioned close to each other. In other words we deal with a Travelling Salesman Problem (TSP)~\cite{tsp}. Distances between labels are calculated based on their vectorial representations in the semantic matrix~\cite{koopman2015_ariadne}. Then we apply one of the standard algorithms for the TSP~\cite{tsp} and implement a simplified version of the Chained Lin-Kernighan heuristic~\cite{chainedtsp}. In the last step, we re-computed the NMI scores between these labels and all the clusters from all the solutions, using Eq.~\ref{eq.nmi}. Eventually each cluster is represented by a 61 dimensional vector, which we call the \textit{fingerprint} of a cluster. We use this vector as global lexical coordinate system, based on which we can now compare individual clusters, within a solution or across solutions, in a more visual way.  

%ASFeb24: I like this text (above) it is clear and good to read. Curious to see if others also think so.

\begin{figure}
\centering
\begin{tabular}{c}
\hspace{-5em} \includegraphics[width=1.2\linewidth]{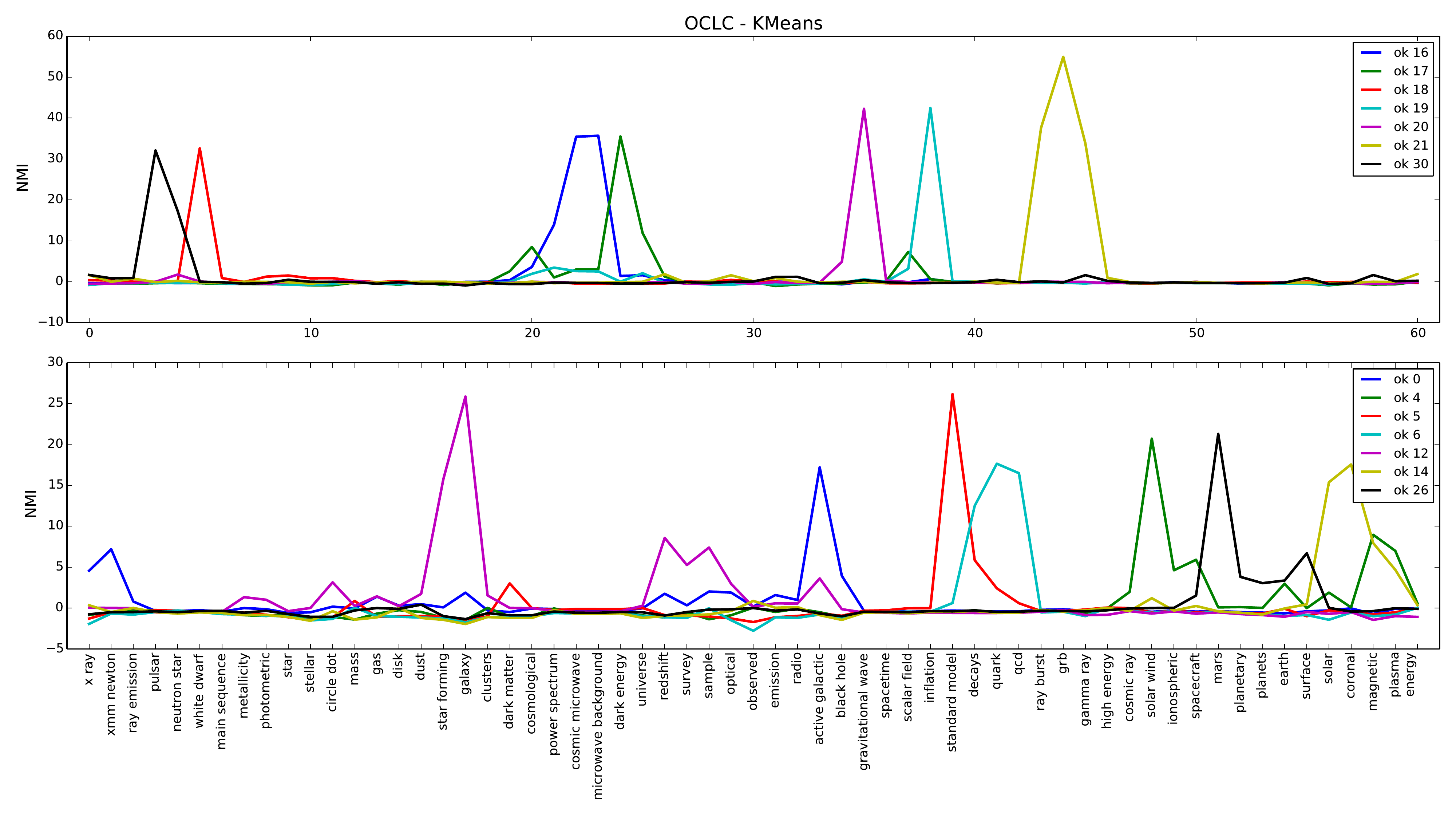}\\
\hspace{-5em} \includegraphics[width=1.2\linewidth]{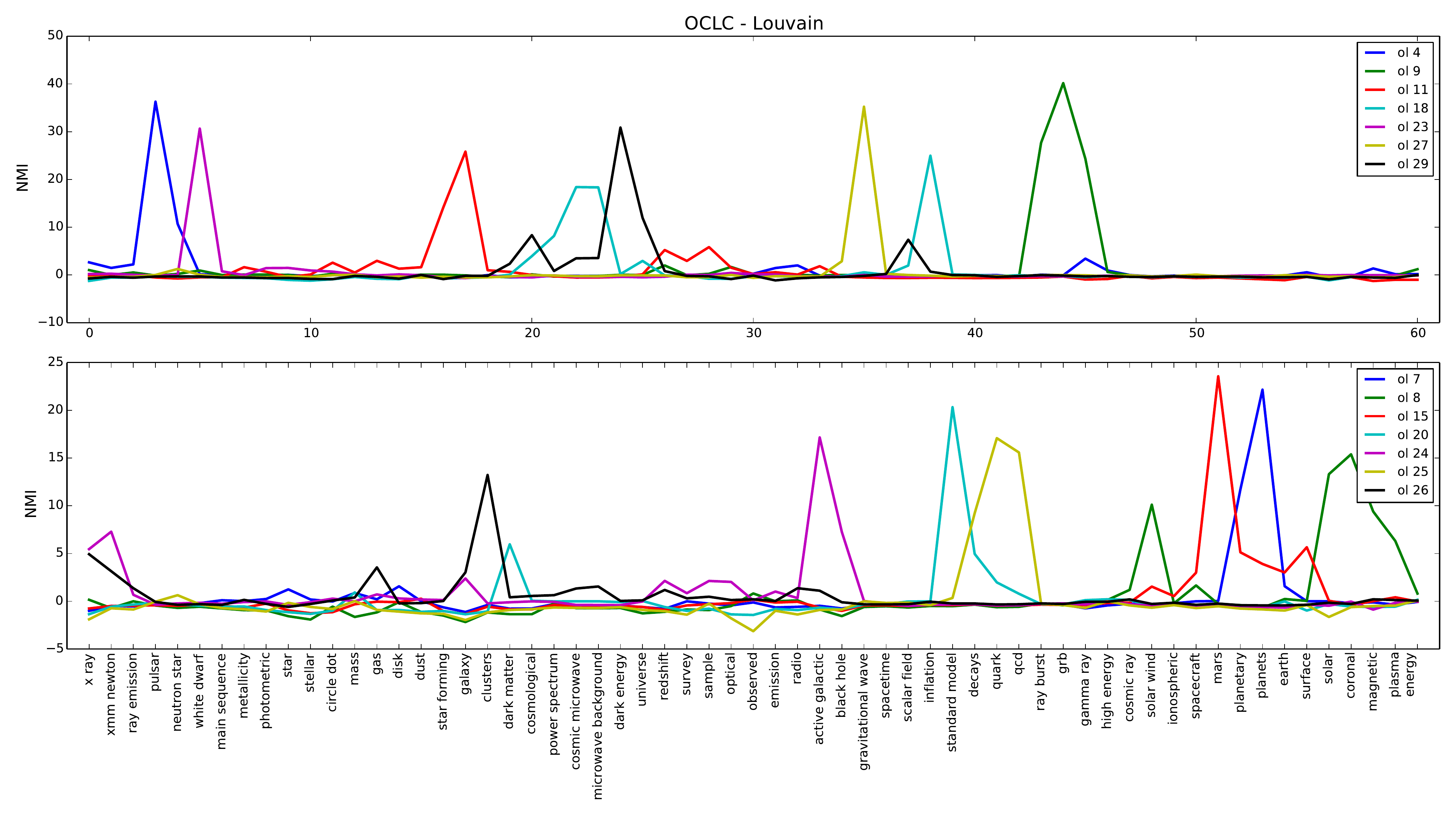}\\
\end{tabular}
\caption{Visual comparison between clusters' labelling fingerprints \label{fig.nmi}}
\end{figure}

In Figure~\ref{fig.nmi}, we visualise selected K-Means and Louvain clusters in terms of their fingerprints. The selection of these clusters is based on the fact that their most informative labels have a very high absolute NMI score. Given the fact that  most of the labels have a very low or even negative NMI values, the high peaks at a very few and sometimes unique labels are very informative about what these clusters are about. For example, we are almost certain that $ok 19$ is about ``inflation,'' $ok 18$ is about ``white dwarf,'' and etc. Actually, the 7 clusters in the upper figure for each solution, for example, $ok 19$, $ok 18$, etc. for K-Means, have more distinguishing labels than those in the figure below, i.e, their highest NMI scores are higher than those in the figure below. We are more certain about the topics of these clusters. The clusters in the lower figures for both solutions have relatively less distinguishing (multiple peaks with lower NMI scores) yet still pretty informative labels. 

\begin{figure}
\centering
\begin{tabular}{c}
\includegraphics[width=.8\linewidth]{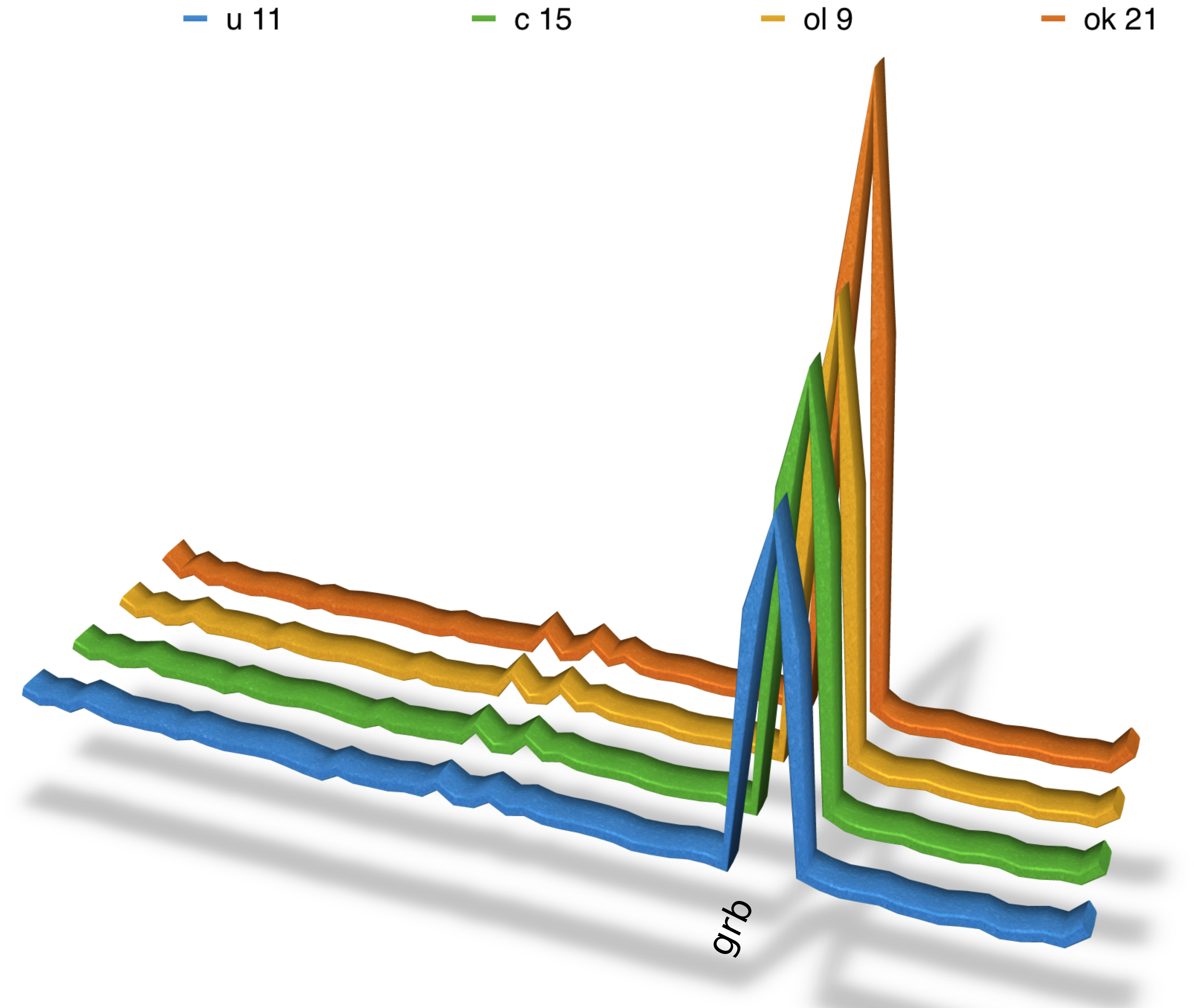}\\
(a) clusters about ``grb''\\
\includegraphics[width=\linewidth]{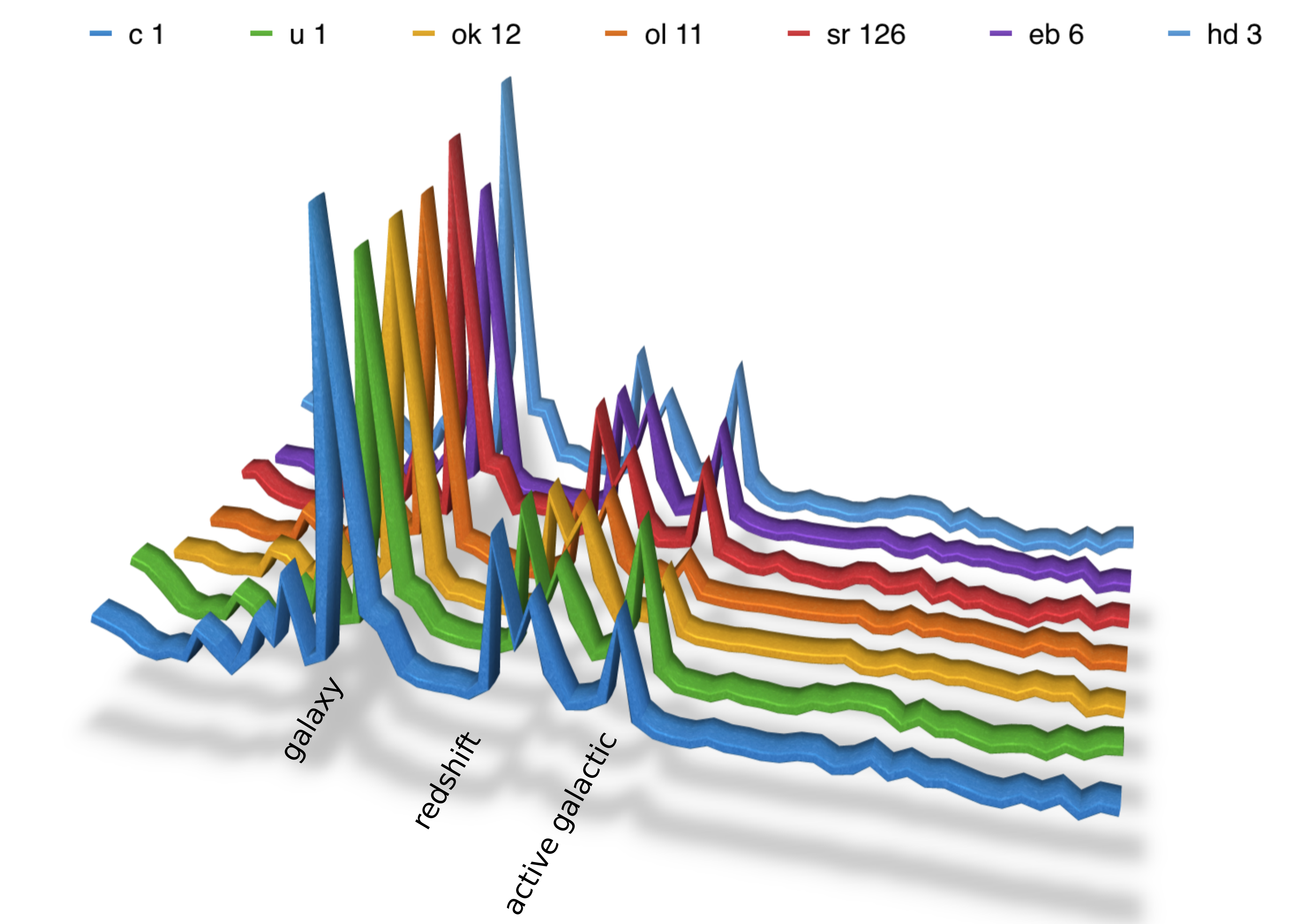}\\
(b) clusters about ``galaxy''\\
\end{tabular}
\caption{Visual comparison between clusters \label{fig.compare}}
\end{figure}

It is also possible to group the individual clusters from different solutions based on their fingerprints. Figure~\ref{fig.compare} shows two groups of clusters. In Figure~\ref{fig.compare} (a), four clusters from four clustering solutions have highly similar fingerprints and they all have one single focus: ``grb'' (gamma ray burst). %\footnote{The  \textit{LittleAriadne}~\cite{koopman2015b} only extracts up to two-word terms from the data, which is why the topic ``gamma ray burst'' is divided into three terms.} (peaks at ``grb,'' ``gamma ray'' and ``ray burst'')
%ASFeb24: Why you thought to take the footnote out again?
Looking at these four clusters more carefully, they share a large overlap in terms of articles they contain. The average Jaccard similarity coefficient\footnote{\url{https://en.wikipedia.org/wiki/Jaccard_index}} among them is 0.64. In Figure~\ref{fig.compare} (b) the seven clusters also have very similar but more complicated fingerprints, with the top three peaks at the ``galaxy,'' ``redshift'' and ``active galactic.'' Their average Jaccard similarity coefficient is 0.47. 

It is not surprising to see that the fingerprints calculated from the NMI scores are consistent with the simple set-based similarity between clusters. 
%ASFeb24 Here you need a reference to the comparison paper, where the set-based simularity is calculated if I'm not mistaken.
If two clusters overlap more, then their fingerprints are more similar. Actually, after applying a simple Affinity Propagation clustering algorithm\footnote{We applied the Python package provided at \url{http://scikit-learn.org/stable/modules/generated/sklearn.cluster.AffinityPropagation.html} with all default parameter settings.
%ASFeb24 The above sentence is not complete: what happened after applying the algorithm?
Further investigation with this clustering exercise is out of the scope of this paper.} over these fingerprints of  clusters from different clustering solutions, there are 22 ``clusters of clusters,'' including the two shown in Figure~\ref{fig.compare}. This may suggest that there exist core articles who have a tendency to always cluster together
%whose clusters are relatively - ASFeb24 check for the change I made
and are recognised by different methods, in other words, they are \textit{prototypical} articles which clearly represent certain topics agreed by different methods, while other articles are in between topics, forming the fuzzy boundaries among topics. Different methods have different ways of deciding the boundaries, which makes the study in this special issue interesting. More comparison between different methods can be found in~\cite{velden2015comparison}.

\section{Conclusion}

In this paper, we showed that using Normalized Mutual Information between clusters and topical terms extracted from titles and abstracts  is an effective way to identify important topical terms to describe clusters. It is a data driven approach which can clearly scale, and has the advantage that the process is independent from clustering methods, and so probably more objective than human judgement. The discussion with a domain expert also showed that these labels represent information he could interact with and which related to his own understanding of the field. The chosen labels are meaningful and useful in follow-up human interpretation and ordering. However, having said this, other labels chosen by other techniques might have a similar function. The aim of this exercise was not to find the \textit{best} labels, but labels which can claim some representativeness and meaningfulness, and labels which enables further semantic interpretation. 

Once a selection of labels is determined to be \textit{lexical reference} or \textit{lexical coordinates}, we can compare different clustering solutions at a global level and also to map single clustering solutions against each other. We showed how such common label-based coordinates can be used to visually compare different clusters or generate ``clusters of clusters''. This way of visual comparison is intuitive and straightforward. It is surprising yet understandable that different clustering methods, despite their differences in data models or algorithms, do share a fair amount of terms which are most informative about their clustering results. The most important result of this paper is a method that uses NMI measures based on lexical information from the documents (articles) which are clustered. We have shown that such a lexical based comparison complements the comparison of clusters in \textit{LittleAriadne}~\cite{koopman2015_ariadne}. Its findings are more or less consistent with other methods of comparison~\cite{velden2015comparison}. 

\section*{Acknowledgement}
Part of this work has been funded by the COST Action TD1210 Knowescape. We would like to thank Marcus John, who functioned as domain expert, for his valuable analysis of the cluster labels. We would like to thank Michael Heinz for discussions on the method.

\bibliographystyle{spmpsci}      % mathematics and physical sciences
\bibliography{astro}   % name your BibTeX data base

\end{document}